# Preprint

# An Overview on Designs and Applications of Context-Aware Automation Systems

Nada Sahlab*, Nasser Jazdi*, Michael Weyrich*

*University of Stuttgart, Institute of Industrial Automation and Software Engineering
Pfaffenwaldring 47, 70569 Stuttgart, Germany. Firstname.lastname@ias.uni-stuttgart.de

**Abstract**

Automation systems are increasingly being used in dynamic and various operating conditions. With higher flexibility demands, they need to promptly respond to surrounding dynamic changes by adapting their operation. Context information collected during runtime can be useful to enhance the system's adaptability. Context-aware systems represent a design paradigm for modeling and applying context in various applications such as decision-making. In order to address context for automation systems, a state-of-the-art assessment of existing approaches is necessary. Thus, the objective of this work is to provide an overview on the design and applications of context and context models for automation systems. A systematic literature review has been conducted, the results of which are represented as a knowledge graph.

*Keywords:* Systematic Literature Review; context-aware automation systems; cyber-physical systems; knowledge graph

## 1. Introduction

Automation systems are continuously evolving and seamlessly integrating in almost all aspects of life. Examples span from the manufacturing domain to surgical rooms, vehicular systems as well the own living space. This vast representation exposes systems to dynamic operating conditions such as varying environmental conditions or dynamic networking and interactions with other systems or users. Therefore, automation systems need to adapt to these conditions by enhancing their decision-making capabilities at runtime [1]. Context-awareness is a system's ability to perceive its operational conditions and relate it to the system operation. Context-awareness has been established in the past decades in the areas of pervasive computing and is mitigating from customizing applications and services based on pre-defined context parameters to a holistic system's ability to support its autonomy and operation under varying surrounding conditions. [2–4] In a recent study, industry practitioners were asked to assess context-aware functionalities in requirements-based systems engineering. The result indicates that including context-aware functionalities is a valuable system feature. However, a clear challenge lies in the lack of experience and standards for designing systems to be context-aware. [5] Several survey papers addressed context from several perspectives, such as on an architectural level, with relation to the Internet of Things and from the perspective of intelligent environments as opposed to computer science. They demonstrate the evolution and extension of the context applicability and scope and show variations in its definition, classification and categorization, modeling as well as reasoning approaches. These surveys give a significant overview, but address projects and contributions, which are outdated considering the fast-paced advancements of the past 5 years in automation systems. Furthermore, the existing literature reviews are only partly focused on the role of context for cyber-physical automation systems.

Paradigms of the Internet of Things (IoT), Cyber-Physical Systems (CPS) as well as the increasing application of Artificial Intelligence (AI) contribute to enhancing the acquisition and processing of embedded and connected heterogeneous data as well as applying remote computing resources to analyze and make predictions. They are both enablers for increased context-awareness and can also benefit from considering context to enhance decision-making and the overall system performance.

For this reason, in this work, we investigate methods and applications of considering context within the operation of automation systems from the past 5 years by presenting the results of a conducted systematic literature review. The findings are represented in the form of a knowledge graph to enable knowledge inference, gaining insights about the methodological developments in the past years as well as making projections about its increasing significance. This contemporary analysis shows the added value of considering context for the systems and classifies applied methods for its modeling and processing.

## 2. Preliminaries

In order to proceed with investigating and analyzing the existing approaches, it is essential to introduce some definitions and concepts related to context-aware systems in the domain of networked automation systems, while also providing a definition of the latter.

### 2.1. Cyber-Physical Automation Systems

Cyber-Physical Systems denote an integration of hardware-based systems with a cyber-representation. [6] The hardware part constitutes of physical sensors, actuators, embedded computing units as well as networking and communication interfaces, while the cyber part includes digital models and further modules supporting the operation of the system. [7] Depending on the granularity and availability of the system's models, a CPS can have a Digital Twin within the cyber-layer when considering further aspects such as the synchronization of the physical and virtual system, relations between virtual models as well as the ability these models to be used for simulations. [8] [9] Due to the networking abilities, CPS can collaborate with other CPS in a dynamic and flexible manner. [10] The communication interfaces allow for using remote computing resources, which supports the increased use of AI to analyze streaming CPS data and influence its operation.

### 2.2. Context-Aware Systems

Context-aware systems are, for the scope of this work, considered as automation systems capable of relating their runtime context to the system's operation. They can therefore acquire context parameters, process, and include them in the decision-making process. This ability can either be an integral part of the system or via remote computing resources with interfaces to the system's operating environment. Some publications include the ability to identify, consider and adapt functionalities based on context as context-awareness, whereas others refer to the latter as context-adaptivity. [3] Context-awareness may optionally lead to an adaptation but can also enable an improved analysis of the system or faults.

Many publications considering context first provide a definition of it. This is due to the fact that defining context is in itself dependent on the context of its application. [11, 12]

Convinced that there is no universally applicable definition, some publications describe context as a relational property providing a better understanding of a situation. [12] In linguistics, context is an implicit property that leads to making alternate or guided conclusions about speech or text. Cognitively, context is an ability in the line of reasoning to extract and transfer historic or present contextual conditions in a present occurrence and use this meta-level information to expand one's own knowledge about historic, present or future occurrences. [13] For software applications and services, context refers to a description of relevant situational parameters, in which the application and service operate. [14] Identifying the relevance of these parameters to the applications as well as detecting variations of them aims at improving the functions delivered to the users of this application in terms of varying the functionality based on variation of the context parameters. Given these definitions and meta-understanding of what constitutes context, a proper description for cyber-physical automation systems can be as follows. Regarding the physical parts of the system, context refers to information from the physical environment of the CPS, which characterize its state and contribute to a better understanding of its operation and analysis of its data in the cyber layer. [15]

## 3. A Systematic Literature Review on Context-Awareness for Automation Systems

In this chapter, the conducted systematic literature review is presented. Starting with briefly addressing available survey papers addressing the same topic, the design characteristics with the research questions and inclusion criteria are then presented, followed by a synthesis of the results and an analysis of the investigated literature, which comprised of 153 publications.

### 3.1. Previous Surveys on Context-Aware Systems

Several surveys capturing state of the art with regard to context already exist; they each have different foci and perspectives. Some of these surveys are briefly described below.

1. *"The practical role of context modelling in the elicitation of context-aware functionalities"*
   In this survey [5], software engineering to support ubiquitous computing is set in focus. The aim of the survey was to show present approaches and identify gaps for considering contextual elements in requirements engineering for practitioners to elicit context-aware functionalities. The scope was for applications and designing a context model, from which functionalities later on can be derived. The study aimed at assessing the effort and knowledge required. For this reason, 37 professionals were interviewed.
   The results show that although context-aware functionalities were valued, context modelling was perceived as complex with a lack of standardized design approaches and domain expertise.

2. *"Context Provisioning Middleware"*
   In [3] from 2013, the focus was on context middleware to encompass the steps and functions necessary for context provisioning. The study highlighted on context classification based on spatial, temporal, device, network and communication, environment, user, activity, among others. The evolution of context and its widening scope was addressed as well. Then, an architectural evaluation of middleware design approaches was presented, for which 16 projects were investigated.
   Context awareness was found to be a key feature for ubiquitous computing. Furthermore, a future projection predicted that the number of context parameters will increase with the advancement of the Internet of things.

3. *"Context Aware Computing for The Internet of Things"*
   [4] investigated literature from 2001-2011, where 50 projects were portrayed. The focus lied in context-aware computing of sensor data. Primary and secondary context were described as categories and a list of context types was provided. The study highlighted on challenges for identifying sensor context and that IoT requires the application of different modelling, acquisition, and reasoning techniques for context.
   The clear importance of context for IoT was also highlighted.

4. *"A Survey on the Evolution of the Notion of Context-Awareness"*
   [13] investigated approaches for context reasoning from 2 different perspectives, namely AI and Intelligent Environments (IE). Differences and synergies between both fields were described. While the focus of AI lies in learning context for knowledge representation and inclusion in the line of reasoning, IE adopts a more pragmatic approach for what can be identified as context by a system and what it enables the system to do.
   The consideration of context for IE should support the transition from techno-centered to human-centered systems, where the technology adapts to humans and not the other way around, elevating systems from simple automation to intelligent ones. Context learning is essential was highlighted to be an essential step for this transition.

5. *"A consolidated view of context for intelligent systems"*
   [11], a study of 2017, 36 context models were viewed that enable systems to adapt their behaviour. Sources of publications were also indicated (ACM, IEEE and Web of Science), snowball search method was conducted, where from 47 publications 36 were extracted. The respective context models from various application fields were then analysed in terms of the elements considered in the model and the variety of

those elements was presented.

User and environmental context parameters were prominent among the investigated models. Lastly, consolidated findings about used context elements were presented.

The surveys studied address different aspects of context. However, one can draw a few general conclusions. The authors would like to present the following relevant conclusions from the investigated papers:

- It is evident that the role of context for enhancing cyber-physical automation systems is significant and is increasingly considered by system designers and industry practitioners.
- An evolution in what constitutes context as well as which methods can be applied to support its acquisition and analysis exists. Context modeling approaches are therefore needed to have an underlying representation of contextual elements and enable a dynamic extension. Hybrid approaches for processing and learning based on the context model should be applied.

## 3.2. Design Preliminaries

A Systematic Literature Review (SLR) is a structured manner for performing surveys on scientific topics by following a predefined plan. It has three main phases, a definition phase, where the problem statement and research questions is defined. Based on the scope, the review is designed with keywords as well as inclusion and exclusion criteria. In the second phase, the review is conducted by searching, filtering and analyzing found results. In the third phase, the results are presented. [16] Our goal is to capture the state of art for making projections about the future direction for context-aware systems in the field of automation technology. Usually, SLRs are conducted manually. In our approach, we designed a mechanism to automate parts of the search as well as represent the results in the form of a connected graph to enable an improved analysis. We adopted the characteristics of an SLR in our conducted survey in the following form:

**Research Questions**

In our survey, we formulated the research questions based on What, How, Why and Where:

- **What**: Referring to the scope of context in a publication, how it is defined and how variable the context parameters are.
- **How**: Referring to the methods used for modeling and reasoning over context.
- **Where**: Application areas within automation technology, for which context is considered.
- **Why**: The added value that context brings in a certain use case.

**Keywords**

For the inserted keywords, some variations were considered to assess a wider scope of publications in the desired domain. A combination of context-awareness and industrial automation system sets the desired search scope. For each of both keywords, some variations are possible, which were considered. These variations are shown in Table 1.

Table 1. Keyword variations

| Keywords | Field of Application |
|---|---|
| [( Context Awareness ) OR | (Automation Systems) OR |
| (Context sensitive) OR (Context adaptive) OR  (Context alone) ] AND | (Production Systems) OR |
| | (Mechatronic Systems) OR |
| | (Control Systems) OR |
| | (Cyber Physical Systems) OR |
| | (Knowledge-Based Assistance Systems) OR |
| | (Autonomous Systems) OR |
| | (Manufacturing Systems) OR |
| | (Smart Factory) OR |

## Inclusion Criteria

Next, the following inclusion criteria were defined for the survey:
- The publications considered should be from the past 5 years.
- The publication language should be English.
- The publication should consider automation systems with no restriction on the domain such as the industrial or daily living one.

## Exclusion Criteria

The exclusion criteria considered for this survey were as follows:
- Work only focusing on context-awareness for service orchestration or recommendation systems without a relation to automation systems is excluded, including work on mobile application development or service design.
- Publications addressing context-aware computing in terms of data analysis, networking or routing without a relevance to an automation system are excluded.
- Survey papers, invited/ featured articles or papers addressing current and future trends without proposing a specific approach for considering context will be taken as reference and benchmark for the conclusions drawn from this SLR but will not be a part of it.

## Databases

Choosing the sources of a literature review is an essential step and influences the scope of analysis, hence the results. Therefore, the choice of the database was based on reviewed publications, either from conferences or journal publications that have been peer-reviewed. The digital libraries considered for the search procedure were the following:
- IEEE Xplore
- ACM Digital Library
- Wiley
- Science Direct
- SpringerLink

The search methodology used the extended search feature to consider the inclusion criteria. The access to the database was possible for IEEE, Wiley as well as SpringerLink, partial access was available on SpringerLink as well as ACM. In these cases, the screening was based on the title, the keywords as well as the abstract. Pre-prints or available copies stored on the author's personal or institutional websites were acquired accordingly.

## 4. Results

For the analysis and results part, this investigation takes a new approach. First, an overview on the application domains, the context modeling and reasoning methods related to each publication are extracted and represented as a knowledge graph. For each publication, some key aspects are shown, such as the name of the publication, its authors, publication year, publisher, URL as well as citation count. The knowledge graph representation approach is discussed in the following section and provides an overview on the applied methods. In the second part of the analysis, selected approaches from literature are chosen, which consider context as a system enabling characteristic and address methods to its dynamic modeling and provisioning at runtime.

### 4.1. Statistics about the literature findings

Based on the design preliminaries and the exclusion and inclusion criteria, a total of 153 publications were considered and further analysed. Figure 1 shows some classifications. Most of the publications were from Springerlink, whereas IEEEXplore came in second place. An initial categorization shows that the industrial automation domain takes the lead, with more emerging publications considering context-awareness in the field of transport and mobility. A possible explanation is that the industrial domain increasingly adopts methods to realize Industry 4.0 concepts whereas considering autonomous vehicles is increasing.

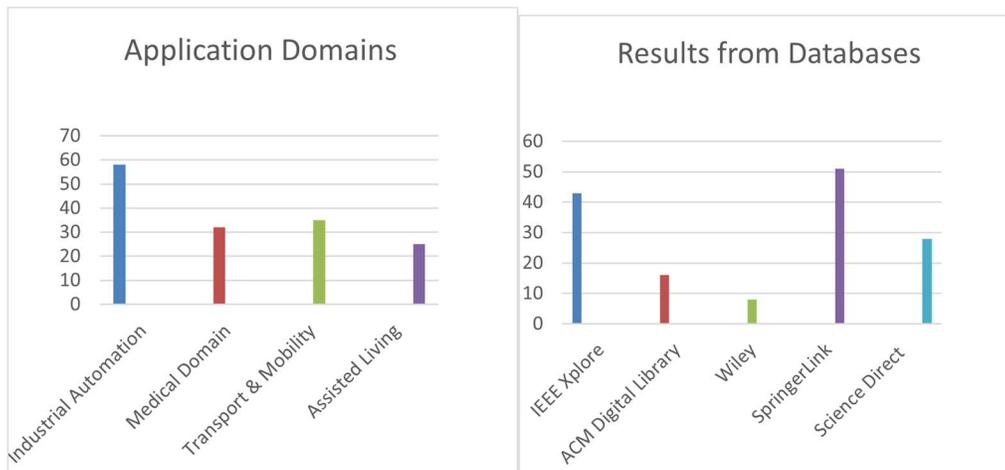

Fig. 1. Classification on filtered literature based on database and application domain

Fig. 1 shows selected use cases, for which context was considered and was expected to bring an added value. Decision-making support takes the lead in those use-cases, whereas use cases such as energy efficiency, process and data management and increasing safety also considered context in significant numbers.

Fig. 3 shows a further correlation of considering and using context with different systems or approaches. Half of the publications addressed context-aware computing, whereas some addressed autonomous or self-aware systems. 13% used context in relation with data collection or processing. Data constituting context mostly came from sensor data, while very few approaches considered further heterogeneous data sources such as documents and further media.

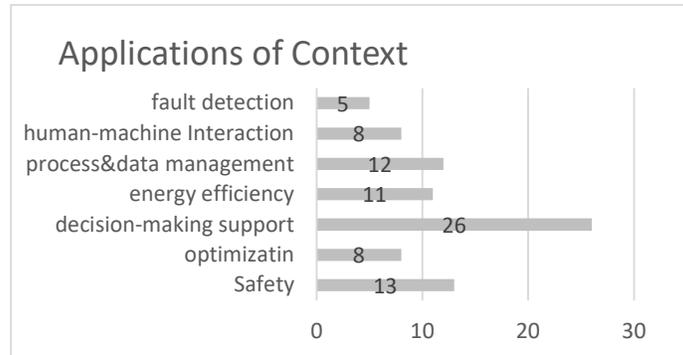

Fig. 2. Classification on filtered literature based on database and application domain

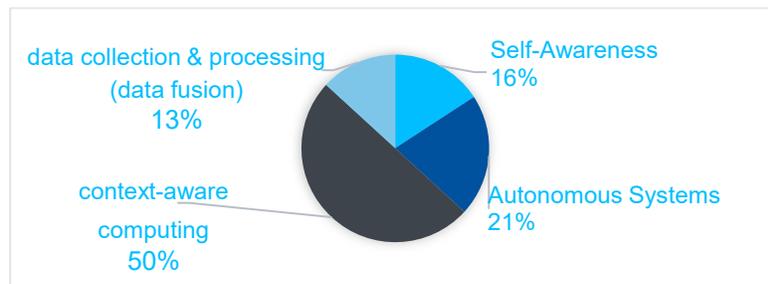

Fig. 3. Classification on filtered literature based on database and application domain

## 4.2. Representation as Knowledge Graph

A knowledge graph is an approach to represent emerging knowledge, while highlighting on the relations between modeled entities. [8] With this representation, deriving conclusions and inferring new knowledge is possible. Based on the SLR criteria introduced, the resulted literature was mapped into a knowledge graph. The aim was to be able to analyze emerging approaches and show possible relations between the methods used for modeling and reasoning on context. The knowledge graph was created as a labeled property graph with an imported CSV file with the extracted results. The number of publications investigated was 153. Each publication node was represented as the metamodel shown in Fig. 4. Using cypher queries, parts of the modeled knowledge graph can be extracted and the relation to other entities shows. For example, when it comes to context modeling, extracted approaches are related to each publication. Fig. 5a shows the mapping of the modeling approaches to the publications. It is evident, that the modeling approaches have some relations to each other and that a publication can have more than one modeling approach.

Predominantly, context is modeled as a graph to highlight on the contextual relations as well. A graph-based approaches comes in relation to the ontological one, for example, an ontology can be represented as an RDF graph.

Key value pairs also come in relations to graph, for example, when modeling context using property graphs with further properties described as key-value pairs.

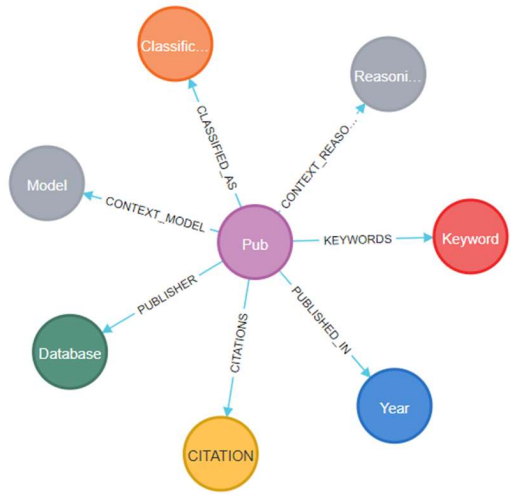

Fig. 4. Structure of the publication node in relation to citation count, classification, database, keywords, year of publication as well as context modeling and reasoning approach

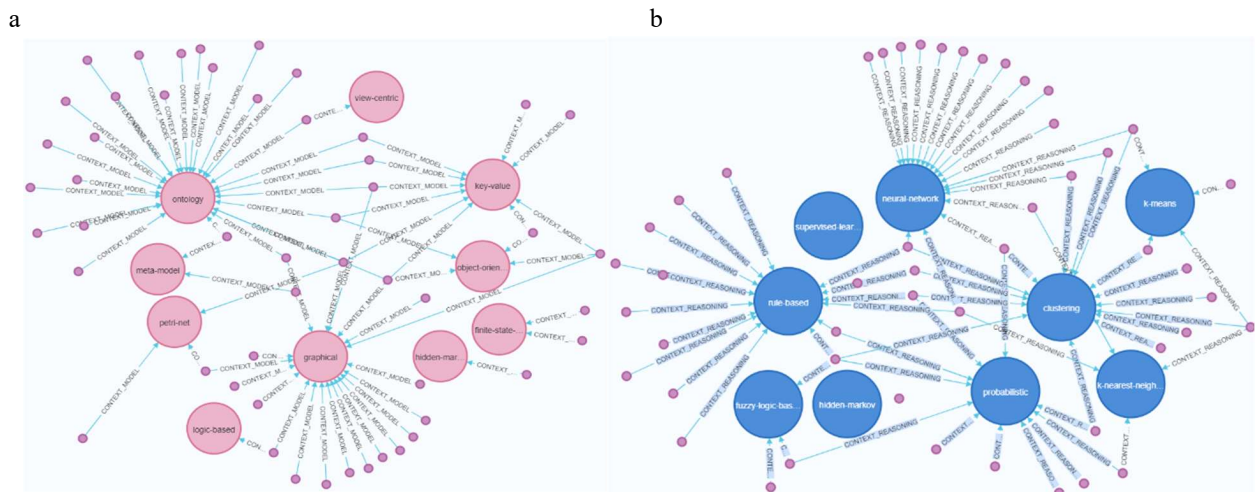

Fig. 5. a: context modeling approaches classified. b: context reasoning approaches classified

Hybrid modeling approaches are emerging as a way to accommodate and manage the increasing complexity of the context parameters and set focus on their relationships to each other and the system. It should be noted that not all publications had a context model.

As for context reasoning and analysis approaches, a query showing this graph extract can be seen on Figure 5b. The publication nodes are mapped to different methods extracted and generalized from the analyzed literature. Here, relations between different approaches are also present, showing the increased employment of hybrid reasoning approaches. Counter to the previous surveys, emerging AI-methods such as clustering and hidden Markov layers are employed, as opposed to rule-based reasoning approaches, which are mostly used in combination with ontologies.

For publications with a context model, reasoning was based on a closed world assumption with a gap for applying methods of unsupervised learning or the extraction of features to improve unsupervised learning techniques. How closely modeling is related or suitable for reasoning is an aspect that still needs to be investigated.

### 4.3. Selected Approaches

The investigated publications mostly used either a context model or considered a few context parameters for data processing while realizing the different use cases. The scope of context parameters is concluded to be limited and static. The focus is therefore set on some prominent approaches that propose structured approaches to consider an evolving context scope at runtime. These approaches are presented below.

- Cameonto [17]: In this approach, a meta ontology was defined with major classes to reflect on common context entities and to enable a transferability across various application domains. For each of the necessary steps for using context such as detection, acquisition and application, services have been defined in a middleware-based architecture.
- In the "The CASAD Matrix Method" [18] an approach to assist system designers in documenting, analyzing and optimizing context-aware systems is introduced. The approach applies a matrix-based method to represent context-related properties and their correlations in a structured manner, thereby reducing the complexity of designing a context-aware system. Noteworthy is that the proposed, domain-agnostic matrix allows for different degrees of detail to be considered and the use of sub-systems
- "Designing context-aware systems" [12]: A method for understanding and analysing context in practice: In this work, a structured approach is proposed that enables the identification of relevant context upon which needed system components to detect and adapt to the it can be determined. The process starts with identifying a problem, then analyzing which context elements can be applied to address the problem. In a following step, needed sensors and actuators to acquire context and adapt the system functionality in a rule-based manner are chosen.
- In this work [10], an approach to identify and analyse the context of a Cyber-Physical System within a dynamic network of CPS by using view-centric context modeling is proposed. The work aims at enabling CPS to be more self-adaptive by detecting their emergent behavior. Based on the system functionality, shared and at runtime emerging functional dependencies between different CPS can be determined and documented as an architectural view using the Unified Modeling Language. The approach is limited to a specific use case and as the authors state, can be embedded into an established context modeling approach and verified by further ongoing use cases.

## 5. Summary and Conclusions

Systematic Literature Reviews aim at investigating current approaches to conclude a research gap or determine a futuristic approach. They represent a significant part of a research activity, based on which new concepts stem.

In this work, an overview on the design, modeling, and reasoning of context for automation systems was provided. The automation systems under consideration were cyber-physical systems, due to their cyber-representation and networking abilities that can foster the design and analysis of context. The SLR was processed in a structured manner and the findings represented in the form of a connected graph.

The main findings can be summarized as follows:

As for the application areas, these conclusions can be made:

- A prominent domain for using context is the industrial domain due to the maturity and readiness of Industry 4.0 and CPS concepts e.g., the Digital Twin. However, a combination of context modeling for the Digital Twin is still not widely addressed in the previous publications and remains a potential for future work.
- The mobility domain is also increasing focus on context-awareness to support system autonomy.
- The health sector still has potential, after overcoming challenges with data privacy, focusing on the main context elements, the patient in a continuously evolving manner.

As for the added value that context brings, these conclusions can be made:

- The added values are mostly placed in the operational phase of the system. Most approaches aim at actively influencing the operation at runtime, decision-making: optimization or reconfiguration. Monitoring and predictive maintenance was also addressed.

- Motivation for considering context dealing with dynamic systems in a network, large amounts of data or the increasing complexity. Other motivations included adaptation to dynamic environments, Increasing system autonomy as well as consideration of flexibility requirements.

- A prominent aspect found was the involvement and consideration of actors which can be operators, patients, or drivers.

- Not all literature included an explicit context model, some of which presented approaches for context-aware computing within the scope of automation systems. Of the publications haven an explicit context model, ontological or graph-based modelling are prominent, which use the semantic expressiveness and represent relations between entities.

- As for reasoning, hybrid approaches combining semantic modelling and machine learning are not yet widely considered and present a further potential for future work.

## References

[1] K. Bellman et al., "Self-aware Cyber-Physical Systems," ACM Trans. Cyber-Phys. Syst., vol. 4, no. 4, pp. 1–26, 2020, doi: 10.1145/3375716.

[2] S. Scholze, J. Barata, and D. Stokic, "Holistic Context-Sensitivity for Run-Time Optimization of Flexible Manufacturing Systems," Sensors (Basel, Switzerland), vol. 17, no. 3, 2017, doi: 10.3390/s17030455.

[3] M. Knappmeyer, S. L. Kiani, E. S. Reetz, N. Baker, and R. Tonjes, "Survey of Context Provisioning Middleware," IEEE Commun. Surv. Tutorials, vol. 15, no. 3, pp. 1492–1519, 2013, doi: 10.1109/SURV.2013.010413.00207.

[4] C. Perera, A. Zaslavsky, P. Christen, and D. Georgakopoulos, "Context Aware Computing for The Internet of Things: A Survey," IEEE Commun. Surv. Tutorials, vol. 16, no. 1, pp. 414–454, 2014, doi: 10.1109/SURV.2013.042313.00197.

[5] R. Falcao, K. Villela, V. Vieira, M. Trapp, and I. L. de Faria, "The practical role of context modeling in the elicitation of context-aware functionalities: a survey," in 2021 IEEE 29th International Requirements Engineering Conference (RE), Notre Dame, IN, USA, Sep. 2021 - Sep. 2021, pp. 35–45.

[6] F. Tao, Q. Qi, L. Wang, and A. Nee, "Digital Twins and Cyber–Physical Systems toward Smart Manufacturing and Industry 4.0: Correlation and Comparison," Engineering, vol. 5, no. 4, pp. 653–661, 2019, doi: 10.1016/j.eng.2019.01.014.

[7] N. Jazdi, "Cyber physical systems in the context of Industry 4.0," in 2014 IEEE International Conference on Automation, Quality and Testing, Robotics: 22 - 24 May 2014, Cluj-Napoca, Romania, Cluj-Napoca, Romania, 2014, pp. 1–4.

[8] N. Sahlab, S. Kamm, T. Muller, N. Jazdi, and M. Weyrich, "Knowledge Graphs as Enhancers of Intelligent Digital Twins," in 2021 4th IEEE International Conference on Industrial Cyber-Physical Systems (ICPS): Online, 10-13 May, 2021, Victoria, BC, Canada, 2021, pp. 19–24.

[9] B. Ashtari Talkhestani et al., "An architecture of an Intelligent Digital Twin in a Cyber-Physical Production System," at - Automatisierungstechnik, vol. 67, no. 9, pp. 762–782, 2019, doi: 10.1515/auto-2019-0039.

[10] B. Tenbergen, M. Daun, P. Aluko Obe, and j. Brings, "View-Centric Context Modeling to Foster the Engineering of Cyber-Physical System Networks," in 2018 IEEE 15th International Conference on Software Architecture: ICSA 2018 : proceedings, Seattle, WA, 2018, pp. 206–20609.

[11] C. Bauer and A. Novotny, "A consolidated view of context for intelligent systems," AIS, vol. 9, no. 4, pp. 377–393, 2017, doi: 10.3233/AIS-170445.

[12] S. van Engelenburg, M. Janssen, and B. Klievink, "Designing context-aware systems: A method for understanding and analysing context in practice," Journal of Logical and Algebraic Methods in Programming, vol. 103, pp. 79–104, 2019, doi: 10.1016/j.jlamp.2018.11.003.

[13] Patrick Brézillon, "Context in Artificial Intelligence II. Key Elements of Contexts," Comput. Artif. Intell., vol. 18, no. 5, 1999.

[14] A. K. Dey, "Understanding and Using Context," Personal and Ubiquitous Computing, vol. 5, no. 1, pp. 4–7, 2001, doi: 10.1007/s007790170019.

[15] N. Sahlab, N. Jazdi and M. Weyrich, "An Approach for Context-Aware Cyber-Physical Automation Systems", IFAC-PapersOnLine, Volume 54, Issue 4, 2021, Pages 171-176, ISSN 2405-8963, https://doi.org/10.1016/j.ifacol.2021.10.029.

[16] Y. Xiao and M. Watson, "Guidance on Conducting a Systematic Literature Review," Journal of Planning Education and Research, vol. 39, no. 1, pp. 93–112, 2019, doi: 10.1177/0739456X17723971.

[17] J. Aguilar, M. Jerez, and T. Rodríguez, "CAMeOnto: Context awareness meta ontology modeling," Applied Computing and Informatics, vol. 14, no. 2, pp. 202–213, 2018, doi: 10.1016/j.aci.2017.08.001.

[18] P. Rosenberger, M. Grafinger, D. Gerhard, M. Hennig, and S. Dumss, "The CASAD Matrix Method: Introduction of a Technique for the Documentation, Analysis, and Optimization of Context-Aware Systems," Procedia CIRP, vol. 91, pp. 427–432, 2020, doi: 10.1016/j.procir.2020.03.106.